\documentclass[aps,prl,twocolumn,a4paper,showpacs]{revtex4}
\usepackage{amssymb,amsmath,graphicx}

\begin{document}

\title{Kinetic roughening in two-phase fluid flow through a random 
Hele-Shaw cell}

\author{Eduard Paun\'e}
\author{Jaume Casademunt}

\affiliation{Departament d'Estructura i Constituents de la
Mat\`eria, Universitat de Barcelona, Avinguda Diagonal, 647, 08028
Barcelona, Spain }

\begin{abstract}
A nonlocal interface equation is derived for two-phase fluid flow,
with arbitrary wettability and viscosity contrast
$c=(\mu_1-\mu_2)/(\mu_1+\mu_2)$, in a model porous medium defined
as a Hele-Shaw cell with random gap $b_0+\delta b$. Fluctuations
of both capillary and viscous pressure are explicitly related to
the microscopic quenched disorder,
yielding conserved, non-conserved and power-law correlated noise
terms. Two length scales are identified that control the possible
scaling regimes and which scale with capillary number as $\ell_1
\sim b_0(c\mathrm{Ca})^{-1/2}$ and
$\ell_2 \sim b_0\mathrm{Ca}^{-1}$.
Exponents for forced fluid invasion are obtained from numerical
simulation and compared with recent experiments.

\end{abstract}

\maketitle

The displacement of a fluid by another in a porous medium is a
problem of fundamental interest in nonequilibrium physics as a
paradigm of interface dynamics in disordered
media~\cite{Barabasi95,Halpin-Krug}. Experiments on bead packs in
Hele-Shaw cells~\cite{RHH} in particular, have
stimulated considerable theoretical efforts, but the problem has
consistently revealed itself rather elusive~\cite{Barabasi95,Halpin-Krug}.
More recently a new
surge of interest has arisen with the recognition of the
inherently non-local character of the problem as a key
ingredient~\cite{Ganesan98,Dube99}, and the realization of a new
series of experiments in Hele-Shaw cells with random
gap~\cite{Hernandez01,Soriano02,Soriano02b}. Roughening exponents of the
proposed nonlocal equations have been explored by means of
Flory-type scaling arguments~\cite{Ganesan98} and phase field
simulations~\cite{Dube99,Hernandez01}. While the specific properties of noise
are known to be crucial to determine the universal aspects of
interface roughening, fluctuations are usually modeled at a
phenomenological level, and including only local capillary
effects. Noise related to the non-Laplacian viscous pressure due
to quenched disorder in the permeability has been so far
neglected. While this may be justified for imbibition
experiments~\cite{Dube99}, other situations, such as forced fluid
invasion, do require a quantitative assessment of this point. In
addition,
it would be desirable to have a unified formulation for general
conditions of viscosity contrast $c=(\mu_1-\mu_2)/(\mu_1+\mu_2)$
and wettability given the rich
variety of phenomena that the experimental evidence has unveiled
as a function of those parameters~\cite{Halpin-Krug}.

Here we address the general problem of fluid displacement in a
Hele-Shaw cell with random gap, as a simple model of a porous
medium. This model system has the great advantage that no
coarse-graining procedure must be invoked in the theoretical
formulation, thus allowing us to derive \it ab initio \rm a
general and complete interface equation, quantitatively accurate,
with explicit dependence on `bare' parameters, and including all
noise sources. On the experimental side, the system is also
appealing since a direct control of the disorder is locally
possible on the microscopic scale~\cite{Hernandez01,Soriano02}.

 A complete
description of interface fluctuations must contain three basic
physical effects of a porous matrix on its motion, namely local
variations of (i) capillary pressure, (ii) permeability, and (iii)
available volume. Different but not independent noise terms must
thus be generated through distinct physical mechanisms from the
unique source of randomness.
Fluctuating capillary pressure can be directly related to gap
variations $\delta b = b-b_0$ where $b_0$ is the mean value, in
terms of the Young-Laplace condition for the pressure jump across
the interface
\begin{equation}
\label{eq:gib-th} p_2 - p_1=\sigma \left (\kappa + \frac{2\cos \theta}{b_0
+\delta b} \right )
\end{equation}
where $\kappa$ is the curvature in the cell plane,
$\sigma$ is surface tension and
$\theta$ is the contact angle, $\cos\theta = 1$ meaning perfect wetting 
of the (invading) fluid 1.
The effect of viscous pressure fluctuations, however, is far less
obvious due to the inherently nonlocal character of the interplay
of (ii) and (iii) in the response of the fluid flow to gap
fluctuations. We base our analysis on the assumption that, for
sufficiently smooth gap variation (i.e. $|\nabla b| \ll 1$)
Darcy's law for a Hele-Shaw cell~\cite{Casademunt92} holds locally as
\begin{equation}
\label{eq:vel}
 {\mathbf v} =-\frac{[b_0 + \delta b(x,y)]^2}{12 \mu} \nabla p.
\end{equation}
In a capillary tube of lateral size $b$ at fixed injection
pressure, from Darcy's law a relative velocity fluctuation scales as 
$\delta v /v
\sim 2\delta b /b$ (larger permeability implies less resistance to
flow) while at fixed flow injection it is exactly the opposite,
$\delta v /v \sim - 2\delta b /b$ (mass conservation slows down
the flow if there is more volume available). In an actual
disordered medium the solution of the whole pressure field will
thus be required to sort out the effective flow conditions at each
location.
A direct consequence of (iii) is that the 2d effective flow, must
give rise in general to a non-conserved interface equation,
precisely to account for volume conservation in 3d. In our case,
3d incompressibility implies that the 2d flow will satisfy
\begin{equation}
\label{eq:cons} \nabla \cdot (b {\bf v})=0,
\end{equation}
where the gap acts effectively as a variable density.

Eqs.~(\ref{eq:vel})~and~(\ref{eq:cons}) imply that the pressure
field is non-Laplacian. In order to obtain a closed bulk equation
which we can project onto the interface degrees of freedom we
treat the pressure perturbatively in $|\nabla b|$. We thus split
the pressure field as $p=p_0+\delta p$ and keep only the order
$|\nabla b|$ in $\delta p$ to obtain
\begin{subequations}
\label{eq:mainaprox}
\begin{eqnarray}
\label{eq:laplace}
\nabla^2 p_0=0 \\
\label{eq:dp} \nabla^2 \delta p +\frac{3\nabla b}{b} \cdot \nabla
p_0 = 0,
\end{eqnarray}
\end{subequations}
where we have neglected higher orders consistently with the fact
that they have also been omitted in assuming local Darcy flow. The
lowest order Eq.~(\ref{eq:laplace}) can be solved as the
unperturbed problem~\cite{Casademunt92,Alvarez01} with the
modified boundary condition Eq.~(\ref{eq:gib-th}) which contains
all capillary effects, while the simplest boundary condition
$\delta p=0$ is then required for the Poisson Equation~(\ref{eq:dp}).
In terms of the Laplacian Green's function $\delta p$ then satisfies
\begin{eqnarray}
\label{eq:bulk} \int_{int} ds {\mathrm G}(x-x(s),y-y(s))
\frac{\partial \delta p}{\partial n} =
\nonumber \\
-\int dx' dy' {\mathrm G}(x-x',y-y')\frac{3\nabla b}{b} \cdot \nabla p_0
\end{eqnarray}
The free-boundary problem
is thus defined by Eq.~(\ref{eq:vel}) specified at the interface,
with $p=p_0+\delta p$, and the boundary conditions at infinity.
Here we focus on the case of forced fluid invasion, where a fixed
velocity $V$ is imposed at infinity and $\mu_1\geq\mu_2$.
We introduce the dimensionless quenched noise as
 $b^2=b^2_0[1 + \xi(x,y)]$.
Noise originated respectively from $\delta b$ in
Eq.~(\ref{eq:gib-th}), from  $\delta b$ in
Eq.~(\ref{eq:vel}) and from $\delta p$ in Eq.~(\ref{eq:dp}) will be called
respectively \it capillary, permeability \rm and \it bulk \rm
noises.

Concerning the scaling properties of the interface, we are
interested in the lowest order approximation on the interface
deviation from planarity, which is relevant in a Renormalization
Group (RG) sense. Our result for the  interface equation in
Fourier space for two-fluid displacement under constant injection
velocity $V$ takes the form
\begin{eqnarray}
\label{eq:hdetk1} \frac{1}{V}\frac{\partial
\hat{h}(k)}{\partial t} = \delta(k) - c|k|\left[1+(\ell_1
k)^2\right]
\hat h(k) + {\cal{N}}_h(k)\nonumber \\
- \frac{1}{2} \left(1+\ell_2 |k|\right) \hat{\xi}_h(k) +
\hat{\Omega}_{LR}(k,t)
\end{eqnarray}
where the lengths $\ell_1$ and $\ell_2$ are defined in terms of
the capillary number $\mathrm{Ca}=12(\mu_1+\mu_2) V/\sigma$
as $\ell_1=b_0(c\mathrm{Ca})^{-1/2}$ and $\ell_2=2 b_0 \cos\theta
\mathrm{Ca}^{-1}$, and where ${\cal{N}}_h(k)$ denotes the leading
(quadratic) nonlinearities
\begin{equation}
\label{eq:nolin} {\cal{N}}_h(k)= -c|k| \!\int_{-\infty}^\infty
\!\!dq [1-{\cal{S}}(kq)]|q| [1+(\ell_1q)^2]\hat{h}(k-q)\hat{h}(q)
\end{equation}
where $\cal{S}$ is the sign function. $\hat{\xi}_h(k)$ is the
Fourier transform of $\xi(x,h(x))$ and the term
$\hat{\Omega}_{LR}(k,t)$ is a long-ranged correlated noise of the
form
\begin{equation}
\label{eq:bulkgeneral}
\hat{\Omega}_{LR}(k,t)=\frac{1}{\mu_1+\mu_2} \left[\mu_1
\hat{\Omega}^{-}(k,t) + \mu_2 \hat{\Omega}^{+}(k,t) \right]
\end{equation}
with
\begin{eqnarray}
\label{eq:split} \hat{\Omega}^{\pm}(k,t)= |k| \int dxdy \,
\xi(x,y+V t)e^{-ikx\mp y|k|}\Theta(\pm y),
\end{eqnarray}
where $\Theta$ is the step function.

Note that the long-ranged term $\hat{\Omega}_{LR}(k,t)$ enters
effectively as an annealed (explicitly time-dependent) noise (see
discussion below).
Eqs.~(\ref{eq:hdetk1}--\ref{eq:split}) constitute our central
result. Note also that in this formulation we have assumed weak noise
so that multiplicative noise terms of order $h\xi$ or nonlinear in
$\xi$ have been neglected \footnote{Terms of this kind coming from
capillary and permeability can be included systematically~\cite{Paune02d}, but 
those from the bulk cannot be obtained explicitly.}.

The linear deterministic part of Eq.~(\ref{eq:hdetk1}) is well
known~\cite{Casademunt92,Alvarez01}. The complete set of deterministic
nonlinearities
can be obtained
systematically using the weakly nonlinear expansion developed in
Ref.~\cite{Alvarez01}. These include the familiar local terms
considered in Ref.~\cite{Ganesan98}, such as $(\nabla h)^2$, but
also nonlocal terms in real space.

The capillary fluctuations give rise to the conserved
(area-preserving) noise term proportional to $|k|\hat{\xi}_h$.
This contribution is associated to the second term in
Eq.~(\ref{eq:gib-th}) and is generated exactly as the usual
capillary term $-|k|^3 \hat{h}(k)$ which comes from the
first term in Eq.~(\ref{eq:gib-th})~\cite{Alvarez01,Casademunt92}.
A conserved noise term of this form was phenomenologically argued in
Ref.~\cite{Dube99}. On the other hand, the non-conserved noise term
proportional to $\hat{\xi}_h$ results from the trivial lowest order
contribution of permeability noise of the form $V\hat{\xi}_h(k)$,
plus a nontrivial local term of opposite sign coming from the 
expression of the bulk noise defined as
$\delta v_\xi \approx
-\frac{b_0^2}{12\mu} \frac{\partial \delta p}{\partial n}$ 
which takes the form
\begin{eqnarray}
\label{eq:delvk} \widehat{\delta v}_\xi (k)= \frac{3V}{2}
\left( - \hat{\xi}_h(k) + \hat{\Omega}_{LR}(k,t) \right).
\end{eqnarray}

We now proceed to sketch the derivation of the bulk noise. For
simplicity we will consider the one-sided case ($c=1$). Our
derivation for $c\neq 1$ follows the formulation of
Ref.~\cite{Casademunt92} but is more involved and will be
presented elsewhere~\cite{Paune02d}. Neglecting orders 
$\xi \partial_x h$, Eq.~(\ref{eq:bulk}) reads
\begin{eqnarray}
\label{eq:bulk1} \frac{-2}{3} \int_{-\infty}^\infty dx'
\ln \left [(x-x')^2 + (h(x) - h(x'))^2\right ]
 \delta v_\xi(x')=
\nonumber \\
\int_{-\infty}^\infty dx' \left \{ \ln \left [(x-x')^2 + (h(x) -
h(x'))^2 \right ] \xi(x',h(x'))
\nonumber \right .\\
\left . +\int_{-\infty}^{h(x')} dy'\frac{2(h(x')-y')}{(x-x')^2 +
(h(x) - y')^2}
\xi(x',y') \right \} \phantom{22}
\end{eqnarray}
where the integral on $y$ of the rhs of Eq.~(\ref{eq:bulk}) has
been integrated by parts and it has been assumed that the noise
vanishes at infinity, $\xi(x,y \rightarrow - \infty) = 0$. It has
also been applied that $\frac{\nabla b}{b} \nabla p_0 \simeq
-\frac{6\mu}{b_0^2} \frac{\partial \xi}{\partial y} V$, and that
the Green function has the form ${\mathrm G}(x,y) =
-(4\pi)^{-1}\ln [x^2 + y^2]$.  With the change of integration
$y=y'-V t$, and to lowest order in the interface deviation $h - V
t$ with respect to the mean interface position~\footnote{Unlike
local noise terms,
the `annealed' approximation is justified in $\delta v_\xi$ as a
leading order, because of its integral form. This means that the
noise acting on a region where the interface is pinned does vary
because the interface is moving elsewhere and it couples through
the bulk.}, Eq.~(\ref{eq:bulk1}) then reads
\begin{widetext}
\begin{equation}
\label{eq:bulk2}
\int_{-\infty}^\infty \!\! dx' \ln |x-x'| \delta v_\xi(x') =
-\frac{3}{2} \int_{-\infty}^\infty \!\!\!dx' \left \{
 \ln |x-x'| \xi(x',h) + \int_{-\infty}^0 \!\! dy
\frac{-y} {(x-x')^2 + y^2} \xi(x',y + V t) \right\}.
\end{equation}
\end{widetext}
Eq.~(\ref{eq:bulk2}) can be explicitly solved
for $\delta v_\xi$ using distribution Fourier 
calculus~\cite{Richards} to yield 
the result Eq.~(\ref{eq:delvk}).

The quenched noise will be typically correlated on a microscopic
scale $a$. For $ka \ll 1$, $\xi$ is effectively white. If
$\langle \xi(x,y) \xi(x',y') \rangle = \Delta \delta(x-x')
\delta(y-y')$ then we have
\begin{equation}
\label{eq:corr} \langle
\hat{\Omega}_{LR}(k,t)\hat{\Omega}_{LR}(k',t')\rangle =
\frac{\Delta}{2\pi}|k|\delta(k+k'){\rm e}^{-|k|V |t-t'|}
\end{equation}
so $\hat{\Omega}_{LR}$ scales as $|k|^{1/2}$ and introduces long-range
memory. Accordingly, low-$k$ behavior is dominated by the local
part of bulk noise. This implies that 3d conservation overcomes
permeability at low-$k$, giving rise to an overall non-conserved
noise with the same sign as the capillary noise.
Although direct computation of both terms in the
bulk noise shows that the local part is typically larger, it is
unclear to what extent neglecting the long-range term may miss
important details of local interface pinning which may eventually
affect the scaling. Furthermore, for $ka \sim 1$, both local and
nonlocal parts of the bulk noise are comparable but then the
annealing approximation may not be justified, and a more careful
analysis of bulk noise based in Eq.~(\ref{eq:bulk1}) may be
necessary~\cite{Paune02d}. The local and non-local contributions
to bulk noise may be of the same order in other situations.
For instance, persistent noise $\xi=\xi(x)$
yields $\delta v_\xi = 0$, with 
non-conserved and conserved noises opposing each other.

One of the salient
features of Eq.~(\ref{eq:hdetk1}) is that the problem has two
characteristic length scales. $\ell_1$ controls the well-known
crossover between (deterministic) capillary and viscous
forces~\cite{Dube99}. The second length $\ell_2$ is a newly identified
one which defines a crossover between conserved and non-conserved
noise. For general viscosities and wetting conditions the
two length scales are arbitrary and define a variety of
possible scaling regimes and crossovers depending on their
relative size. Experiments of a wetting fluid invading an inviscid
one ($c=1$, $\cos\theta=1$) typically have $\ell_2 \gg \ell_1$.
Concerning the nonlinearities, power counting arguments show that
quadratic terms are relevant in the RG sense only if the viscous
term $c|k|$ is absent, that is, for $|k|\ell_1 \gg 1$. This
capillary-dominated regime will always be observed at large $k$
and short times provided $\ell_2 \gg \ell_1$. A crossover to the
viscous-dominated regime will eventually occur for $|k|\ell_1 \sim
1$ and then the quadratic nonlinearities do become irrelevant. A
second crossover within the viscous regime, from conserved to
non-conserved noise, will occur for $|k|\ell_2 \sim 1$. The
explicit knowledge of the bare coefficients of the nonlinear terms
is helpful to assess the validity of the linear approximation.
Numerical simulation of typical cases show that the effect
of nonlinearities in the capillary regime is not appreciable in
reasonable simulation times. The capillary regime, which is
relevant to experiments~\cite{Soriano02b}, is thus well
described by
\begin{equation}
\label{eq:zerocon} \frac{1}{V}\frac{\partial
\hat{h}(k)}{\partial t} = \delta(k) - |k| \left[
(\tilde{\ell_1}k)^2 \hat h(k) + \frac{1}{2} \ell_2 \hat{\xi}_h(k)
\right]
\end{equation}
with $\tilde{\ell_1}= b_0\mathrm{Ca}^{-1/2}$. Since the
quadratic nonlinearities vanish for $c=0$ (for symmetry), we may refer to the
growth described by Eq.~(\ref{eq:zerocon}),
as the universality class of `symmetric fluid invasion' (SFI)~\footnote{With
respect to RG flow of $c$ this is a saddle fixed point.}.

We will now study the SFI scaling by numerical simulation of 
Eq.~(\ref{eq:zerocon}) and also the subsequent crossovers.
The quantities of interest concerning the scaling properties are
the root mean square of the interface height fluctuations, $W$,
and the structure factor $S(k,t)$. The width grows with time as
$W(t) \sim t^\beta$ before saturation, and the saturation width
scales with system size $L$ as $W_{sat} \sim L^\alpha$. The
roughness exponent $\alpha$ can also be obtained from the relation
$S(k,t) \sim k^{-1-2\alpha}$ for long times. For further
reference, we provide the scaling exponents of
Eq.~(\ref{eq:zerocon}) for two cases that are exactly
solvable~\cite{Paune02d}, namely, for persistent noise $\xi =
\xi(x)$, and annealed noise $\xi = \xi(x,t)$ which, assuming 
$\delta$-correlations, give respectively $\alpha=3/2,\;\beta = 1/2$
and $\alpha=0,\;\beta=0$.

\begin{figure}[p]
\includegraphics[clip]{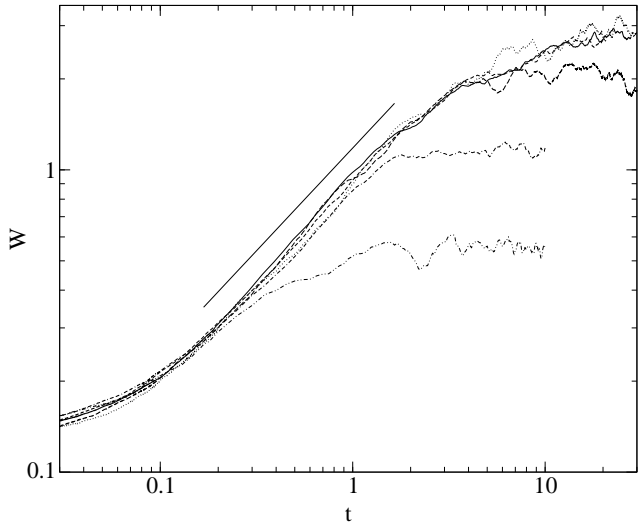}
\caption{\label{fig:W3} Interface width $W$ as a
function of time, for system sizes $L=32$, 64, 128, 256, 512 and 1024.
The straight line is a fit with a slope $\beta=0.68$.}
\end{figure}

\begin{figure}[p]
\includegraphics[clip]{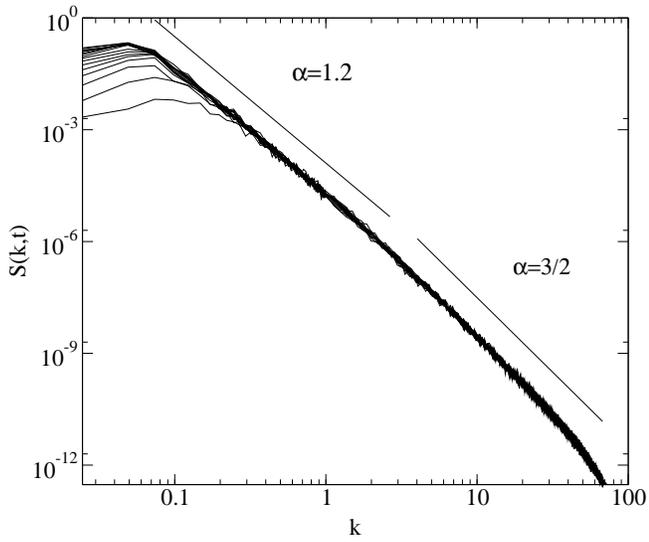}
\caption{\label{fig:S3} Structure factor for a system with $L=256$.
The data are for $t=0.5$ (lower curve)
to $t=12.0$, and time interval $\Delta t = 0.5$. The
straight line with slope $-3.3$ ($\alpha = 1.2$) is a fit, and the
other straight line has slope $-4$ ($\alpha = 3/2$).}
\end{figure}

Typical results for Eq.~(\ref{eq:zerocon}) are shown in 
Figs.~\ref{fig:W3} and \ref{fig:S3}.
We choose $V=1$, $a=0.0625$, $\ell_1=50$ and $\ell_2=3000$, which satisfy
the criterion $a \ll \ell_1 \ll \ell_2$. 
For these values, we have obtained a
roughness exponent $\alpha = 1.2 \pm 0.05$ (very close to the
value reported in Ref.~\cite{Dube99}) and a growth exponent $\beta
= 0.68 \pm 0.02$. The scaling of the correlation function $G(l,t)$
has been found to be fully consistent with
$\alpha_\mathrm{loc}=1$, as expected from the superrough value of
$\alpha > 1$. Note that the scaling of the power spectrum $S(k,t)$
at large $k$ corresponds to the case of persistent noise,
$\alpha=3/2$: short segments of the interface `feel' effectively a
persistent noise for long enough time intervals.

An increase (by an order of magnitude) in the value of $V$ 
with the subsequent variation of $\ell_1$ and $\ell_2$ modifies
the scaling behavior: for low values of $k$ the scaling of
$S(k,t)$ yields $\alpha = 0$, the value obtained for annealed
noise, while for larger $k$ the observed scaling is essentially
the same as in Fig.~\ref{fig:S3}. Hence, the effectively noise acting
on the interface at low-$k$ for large $V$ is annealed. The value
of the exponent $\beta$ is observed to decrease with increasing
$V$, consistently with the trend towards an effective annealed
noise, for which $\beta = 0$. On the other hand, if $V$ is
decreased by an order of magnitude, the scaling observed in 
Figs.~\ref{fig:W3} and
\ref{fig:S3} is essentially unchanged.

The values of $\mathrm{Ca}$ and $V$ we have used are of
similar magnitude to the ones of the experimental work of
Ref.~\cite{Soriano02b}, and the exponents reported there for large
injection rates, $\alpha \simeq 0$ for small $k$ and $\alpha
\simeq 1.3$ for large $k$ are fully consistent with our results
for large $V$. However, this is not the case for small values of
$V$. This discrepancy could be attributed to non-darcyan effects
associated to the sharp edges in the gap variations of the
experiment, which may modify pinning properties.

After the first crossover to the viscous regime with conserved
noise, we find that the interface is not rough, 
$W_{sat}(L \rightarrow \infty) = const$. After the second
crossover to non-conserved noise, we get $\alpha = \beta = 0$,
with $W$ growing logarithmically both with time $t$ and system
size $L$.

As a final remark let us mention that the case of spontaneous
imbibition (constant pressure conditions) can be described with
our formalism with an appropriate time-dependence $V(t) \sim
t^{-1/2}$, except close to pinning, when the zero-mode
fluctuations must be carefully worked out.

We expect that the physical insights and the predictive power of
the theoretical framework here presented may be useful to
reinterpret data and design new experiments, in particular in the
yet unexplored range of parameters. The extent to which
it may be applicable to more realistic porous media, with
appropriate coarse-grained parameters, however, is an open
question which deserves further study. 

We are grateful to O. Camp\`as for helpful discussions.
Financial support from DGES
(Spain) under project BXX2000-0638-C02-02 is acknowledged.

\end{document}